\documentclass[notitlepage, superscriptaddress, 11pt, two-column, single-spaced]{revtex4-1}
\usepackage{graphicx}
\usepackage{subfigure}
\usepackage{amsmath}
\usepackage{epsfig}
\usepackage{epstopdf}
\usepackage{float}
\usepackage{placeins}
\usepackage{amssymb}

\usepackage{rotating}

\usepackage{multirow}
\begin{document}

\title{Breaking new ground in mapping human settlements from space - The Global Urban Footprint}

\author{Thomas Esch}
\affiliation{German Aerospace Center (DLR), German Remote Sensing Data Center (DFD), Oberpfaffenhofen, D-82234 Wessling, Germany}
\affiliation{Chair for Remote Sensing, Institute for Geography and Geology, University of Wuerzburg, Am Hubland, D-97074 Wuerzburg, Germany}

\author{Wieke Heldens}
\affiliation{German Aerospace Center (DLR), German Remote Sensing Data Center (DFD), Oberpfaffenhofen, D-82234 Wessling, Germany}
\author{Andreas Hirner}
\affiliation{German Aerospace Center (DLR), German Remote Sensing Data Center (DFD), Oberpfaffenhofen, D-82234 Wessling, Germany}

\author{Manfred Keil}
\affiliation{German Aerospace Center (DLR), German Remote Sensing Data Center (DFD), Oberpfaffenhofen, D-82234 Wessling, Germany}

\author{Mattia Marconcini}
\affiliation{German Aerospace Center (DLR), German Remote Sensing Data Center (DFD), Oberpfaffenhofen, D-82234 Wessling, Germany}

\author{Achim Roth}
\affiliation{German Aerospace Center (DLR), German Remote Sensing Data Center (DFD), Oberpfaffenhofen, D-82234 Wessling, Germany}

\author{Julian Zeidler}
\affiliation{German Aerospace Center (DLR), German Remote Sensing Data Center (DFD), Oberpfaffenhofen, D-82234 Wessling, Germany}

\author{Stefan Dech}
\affiliation{German Aerospace Center (DLR), German Remote Sensing Data Center (DFD), Oberpfaffenhofen, D-82234 Wessling, Germany}
\affiliation{Chair for Remote Sensing, Institute for Geography and Geology, University of Wuerzburg, Am Hubland, D-97074 Wuerzburg, Germany}

\author{Emanuele Strano}
\affiliation{Department of Civil and Environmental Engineering, Massachusetts Institute of Technology (MIT), Cambridge, MA 02139, USA}
\affiliation{German Aerospace Center (DLR), German Remote Sensing Data Center (DFD), Oberpfaffenhofen, D-82234 Wessling, Germany}

\date{\today}

\maketitle

Today, approximately 7.2 billion people inhabit the Earth and by 2050 this number will have risen to around nine billion, of which about 70 percent  will be living in cities. The population growth and the related global urbanization pose one of the major challenges to a sustainable future. Hence, it is essential to understand drivers, dynamics, and impacts of the human settlements development. A key component in this context is the availability of an up-to-date and spatially consistent map of the location and distribution of human settlements. It is here that the Global Urban Footprint (GUF) raster map can make a valuable contribution. The new global GUF binary settlement mask shows a so far unprecedented spatial resolution of 0.4 arcsec ($\sim12 m$) that provides - for the first time - a complete picture of the entirety of urban and rural settlements. The GUF has been derived by means of a fully automated processing framework - the Urban Footprint Processor (UFP) - that was used to analyze a global coverage of more than 180,000 TanDEM-X and TerraSAR-X radar images with 3m ground resolution collected in 2011-2012. The UFP consists of five main technical modules for data management, feature extraction, unsupervised classification, mosaicking and post-editing. Various quality assessment studies to determine the absolute GUF accuracy based on ground truth data on the one hand and the relative accuracies compared to established settlements maps on the other hand, clearly indicate the added value of the new global GUF layer, in particular with respect to the representation of rural settlement patterns. The Kappa coefficient of agreement compared to absolute ground truth data, for instance, shows GUF accuracies which are frequently twice as high as those of established low resolution maps. Generally, the GUF layer achieves an overall absolute accuracy of about 85\%, with observed minima around 65\% and maxima around 98 \%. The GUF will be provided open and free for any scientific use in the full resolution and for any non-profit (but also non-scientific) use in a generalized version of 2.8 arcsec ($\sim84m$). Therewith, the new GUF layer can be expected to break new ground with respect to the analysis of global urbanization and peri-urbanization patterns, population estimation, vulnerability assessment, or the modeling of diseases and phenomena of global change in general.

\section{Introduction}

Global urbanization represents one of the most urgent present and future challenges. However, the real dimension of this phenomenon is still not completely understood - in particular with respect to a globally precise information basis on the location and distribution of human settlements in urban and - more specifically - in rural areas. Hence, this paper introduces a new inventory of human presence on Earth in form of the Global Urban Footprint raster map that reflects the human settlements pattern in a so far unique spatial resolution of 12m. One of the most frequently referenced figures describing human settlements development is a simple graph composed of two lines representing the urban and rural percentages of the global population between 1950 and 2050. In 1950, the global rural population was twice as large as the urban population. Since then the rural population has been constantly decreasing, while the urban population has drastically increased. By around 2008 the urban population has exceeded the rural one for the first time in human history. This observation, commonly known as the global urban transition (\cite{UNsocial2015}), indicates that the majority of people on Earth inhabit some kind of urban environment. Arguments regarding the magnitude of global urbanization(\cite{united2001,UN20045, UN2006,Birch2011})  have made clear to the global scientific and policy-making community that cities play a primary role as drivers of all social, economic and environmental systems. Currently, there is considerable evidence that global urbanization impacts the entire spectrum of human and natural systems, in particular with respect to energy, water, food, biodiversity, climate  or human health (\cite{Moore2003,Zhou2004,kaufmann2007,Grimm756,Tilman2011}). The benefits and challenges of urbanization are comprehensively discussed in (\cite{Daily1992}, \cite{Johnson2001},\cite{cieslewicz2002},\cite{Dye766}, and \cite{seto2011}).
Beyond the global urbanization mantra, which is largely used to promote research agendas focusing on cities, fundamental questions still pertain to the knowledge about the spatial dimension of urbanization: Which proportion of the global land surface is covered with built-up area? What is the ratio between urban and rural settlements area? How many cities are on Earth? Many studies in fact share the view that estimations of the effects of human presence on Earth are strongly biased  (\cite{Potere2007,Potere2009}) and that the dynamics of growth and its economic and social effects are poorly understood (\cite{Batty769}). 
Thereby the weak points regarding an objective view on global urbanization are clear: first, a shared definition of urban opposed to rural areas is missing. This is, in turn, reflected by a second critical issue - biased demography inventories on urban and rural population. Finally, a spatially detailed and up-to-date inventory of the entirety of urban and rural settlements on Earth does not exist so that any analysis on the global extent and development of human settlements is inherent to a certain bias. 
Meeting the third challenge regarding the lack of a detailed inventory of human settlements on Earth was the motivation for the GUF initiative and product, respectively. Earth observation (EO) imagery certainly represents an effective approach to overcome the lack of objective spatial information on the structure and spatiotemporal development of human settlements on Earth (\cite{esch2010a}). The global classification of human settlements is a very specific topic in urban remote sensing because of the necessary trade-off between spatial resolution of the available EO data and ability to collect a global coverage within a reasonable period of time. A comprehensive overview of the available EO-based and EO-supported global geo-information layers on human settlements is provided by \cite{Potere2009} and \cite{gamba2009} and \cite{Ban20151} . As they report, the majority of these data sets are generated from medium resolution (MR) optical EO data, as for instance the largely-established MODIS 500 (\cite{Schneider20101733})  and GlobCover 2009 (\cite{bontemps2011})  land-cover maps with a spatial resolution of around $500$ m and around $300$ m, respectively. Although representing an important source of information, given their low resolution coupled with the existence of different definitions of urban areas, this first generation of global urban mapping products suffered from inaccuracy and considerable disagreements. For example, the results of a simple estimation of the total urban extent vary by an order of magnitude between VMAP0 and GRUMP, from 0.3 - 3.4 million $km^2$. Potere et al. report three main factors that cause such large inter-map differences: i) the varying dates of the map production, ii) different spatial resolution of the underlying data, and iii) the varying definitions of urban land use (\cite{Potere2009}). Moreover, their capabilities to accurately detect and delineate small and scattered villages and towns are quite limited. More recent initiatives aim to provide spatially more accurate human settlements layers based on high resolution (HR) EO data. NASA, for example, released in 2013 a new global night-time light product derived from imagery of the Visible Infrared Imaging Radiometer Suite (VIIRS) on board of the Suomi NPP satellite (\cite{NASA2012}). The European Join Research Center (JRC) with the Global Human Settlements Layer (GHSL) presented a procedure for an automatic extraction of built-up areas by analyzing global Landsat coverage for several time steps (\cite{Pesaresi2016}). Wieland and Pittore also proposed a method based on object based analysis  and SVM to classify urban large areas from Landsat 8 (\cite{Wieland2016294}) .  Miyazaki et al. propose a method based on the integrated analysis of ASTER satellite images and GIS data to produce a new global HR settlement mask (\cite{Miyazaki}). By means of Envisat-ASAR radar imagery, Gamba et al. and Ban et al.  derived a built-up area layer that aims at improving the GlobCover 2009 urban class (\cite{gamba2012,Ban201528}). 
In this publication we introduce a novel global settlements mask in a so far unique spatial resolution, the Global Urban Footprint (GUF), along with a description of the methodological framework deployed to generate this layer. Section 2 includes information about the underlying EO radar imagery collected in the TanDEM-X mission and a description of the Urban Footprint Processor (UFP) suite used to produce the GUF data set. After this technical section we focus on the characteristics of the new GUF product by providing a product specification, presenting the results of a quality assessment, giving a qualitative description of the main GUF characteristics, and showing first results regarding the global settlements statistics derived from the new GUF data. Finally, we draw the conclusions and provide an outlook on the future activities to support global urban observation and management with the GUF and planned add-on services and products.

\section{The Global Urban Footprint framework}
\label{II}

In 2007 and 2010 the German Aerospace Center (DLR) launched the EO radar satellites TerraSAR-X and TanDEM-X, respectively. In particular the ability to collect a global coverage of very high resolution SAR imagery within a comparably short period of time predestined the two missions to support global environmental monitoring activities. In previous studies we demonstrated the potential to delineate human settlements and basic land cover types based on SAR images by performing a combined analysis of speckle statistics and intensity information (\cite{esch2010,Esch2011} ). Encouraged by the promising outcomes of these studies, DLR's German Remote Sensing Data Center (DFD) started the internal Global Urban Footprint initiative. The goal of this activity was the development of a fully-automated processing framework to produce a world-wide map of human settlements in a so far unique spatial detail by analyzing a global coverage of TerraSAR-X and TanDEM-X images collected in the context of the TanDEM-X mission (\cite{Esch2012}). 

\begin{figure}[!]
\begin{center}
\includegraphics[width=1\textwidth]{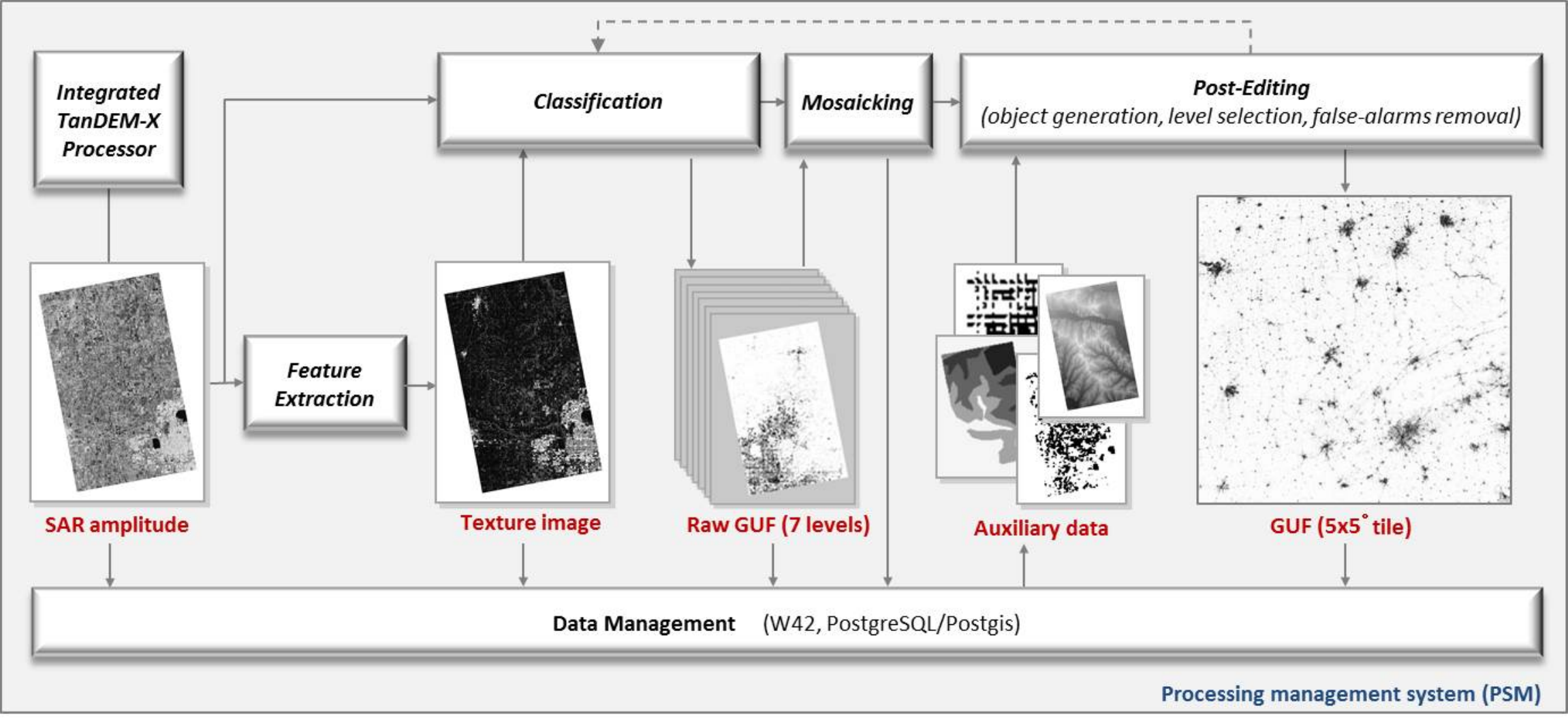}
\caption {Basic scheme of the production framework implemented to generate the Global Urban Footprint layer.}
\label{fig1}
\end{center}
\end{figure}

\subsection{TanDEM-X mission}
In June 2007, DLR launched its first EO radar satellite TerraSAR-X (TSX), followed by the identically constructed TanDEM-X (TDX) satellite three years later (\cite{Werninghaus,Krieger}). Both systems are part of the TanDEM-X mission which is the first bi-static, spaceborne SAR mission. The primary mission goal is the generation of a consistent global digital elevation model (DEM) with an unprecedented accuracy. It furthermore provides a highly reconfigurable platform for testing and demonstrating new SAR techniques and potential applications. Until mid-2014 TSX and TDX were flying in a unique formation with a typical distance between 250 - 500 m. The data acquired in this period form the basis for the global DEM. Since then the orbit configuration was changed several times to enable other experiments.
The required accuracy of the global DEM follows the HRTI-3 specifications. The elevation product is provided in geographic coordinates. The spacing is 0.4” in latitude and longitude which is equivalent to a grid spacing of $12$ x $12$m on the ground. The longitude spacing is varied every $10^{\circ}$ beginning at $50^{\circ}$ and every $5^{\circ}$ from $80^{\circ}$ - $90^{\circ}$ in order to compensate for the convergence of the meridians at the poles. The coarsest spacing is $4^{\prime \prime}$ between $85^{\circ}$ - $90^{\circ}$ latitude. The corresponding SAR images are acquired in single-polarized Stripmap (SM) mode with a resolution of 3.3 m in azimuth and 1.2 m in range. The latter converts into 3.0 - 3.5 m ground resolution depending on the incidence angle selected. The position of the strip acquired during the second acquisition phase (2012) was shifted by approximately $50\%$ compared to the first year (2011). Approximately 470.000 pairs of complex SAR images were processed. Until March 2016 all DEMs for $89\%$ of the Earth's land mass were completed. The DEM production was finished by mid-2016. 
The TanDEM-X mission supports further experiments such as the so-called dual receive antenna mode or polarimetric interferometry, which will be used for the design of future EO SAR missions. A comprehensive description of the potential applications, products and data access procedures are provided in the TDX Science Plan (\cite{DLR1}).
The TanDEM-X mission has been realized in form of a Public Private Partnership (PPP) between DLR and Airbus DS. As a result, the commercial use of products derived from SAR imagery collected in the context of the TanDEM-X mission – hence, including the GUF - is subject to a licensing by Airbus DS, while any scientific use is subject to licensing defined by DLR. The implications of the PPP on the use of the GUF product are further detailed in section 3.

\subsection{Urban Footprint Processor}
The production of the GUF layer is based on the fully-automatic, generic and autonomous processing environment, orchestrating an extensive suite of processing and analysis modules – the Urban Footprint Processor (UFP). This system ensures an effective processing of the 182,249 TerraSAR-X and TanDEM-X single look complex (SSC) image products mostly collected in 2011 and 2012 ($93\%$) in Stripmap mode with 3m ground resolution. To fill data gaps, single scenes collected in 2013 and 2014 were included as well. The volume of this input data set already adds up to 308 TB. Considering all auxiliary data used in the GUF processing, the UFP framework has to handle $>20$ million files with a total volume of $>400$ TB.
The systems design and implementation of the UFP was first described by \cite{esch2013}. However, since then the UFP suite has been optimized and modified, in particular by adding a highly automated post-editing procedure aiming at an elimination of false alarms due to over-classifications. Figure \ref{fig1} provides a schematic overview on the UFP processing environment. Basically, the UFP consists of five main technical modules covering functionalities for data management, feature extraction, unsupervised classification, mosaicking and automatic post-editing.
The UFP is implemented at DLR-DFD as a fully operational, data-driven system deployed on two basic processing platforms, a Sun X 4640 machine with eight CPUs at 2.6 GHz, six cores and 256 GB RAM used for the single file Processing, and a Calvalus Cluster (\cite{fomferra2012}) based on Apache Hadoop consisting of 28 compute nodes with 32 GB RAM and one quad-core Intel Xeon 3.4 GHz CPU  each used for the post-editing steps.

\subsubsection{Data management}
The data management component of the UFP system comprises a set of Solaris servers and storage units with restricted access. In order to manage and orchestrate the work flow, DLR's Processing System Management (PSM) tool was utilized \cite{boettcher2001}. The PSM Tool allows users to integrate any image processing modules, define work flows and issue production requests for each file in a consistent manner. It automatically manages hardware resources and distributes the workload in an efficient way. The job scheduler also allows re-running jobs if any process failed during the production request.
In addition the UFP uses DLR's W42 Raster Data Repository for optimized access to DEM data required in the GUF analysis process. An important feature of the W42 system is its ability to provide a best-of-DEM of a given area from different sources such as SRTM or ASTER (\cite{habermeyer2009w42}).

\subsubsection{Feature extraction}
\begin{figure}[!]
\begin{center}
\includegraphics[width=1\textwidth]{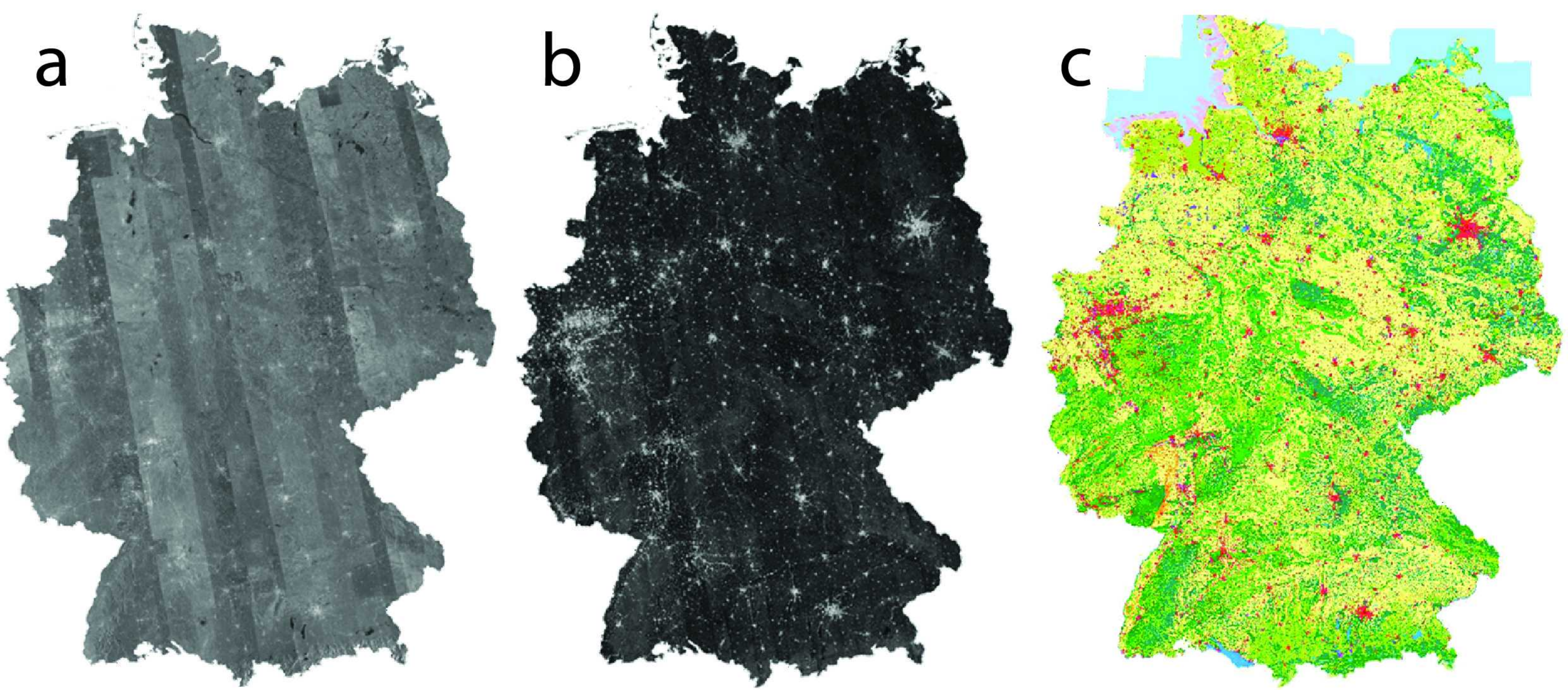}
\caption {Comparison of original TSX Stripmap backscattering amplitude (a), texture feature derived from the amplitude data (b) and CORINE land use data (c).}
\label{fig2}
\end{center}
\end{figure}

\cite{Esch2011} demonstrated that built-up areas show a characteristic small-scale heterogeneity of local backscatter in TSX Stripmap images that can efficiently be used to delineate built-up areas. This effect is related to the specific properties of SAR data for built environments that exhibit strong scattering due to double bounce effects and direct backscattering from the vertical structures (e.g., buildings, bridges, traffic signs). At the same time shadow effects occur on those sides of vertical structures that are facing away from the incoming radar beams. To define this local image heterogeneity, or texture, the UFP calculates the speckle divergence feature defined as the ratio between the local standard deviation and local mean of the backscatter computed in a defined local neighborhood. A detailed description of the feature extraction algorithm is provided in \cite{Each2013b}. 
Figure \ref{fig2} juxtaposes TSX Stripmap amplitude data and the speckle divergence texture image derived from the amplitude data for the area of Germany to a CORINE land cover map . In a visual comparison it is possible to see the strong correlation between the high texture regions appearing as bright spots in the speckle divergence image (Fig. \ref{fig2}b) and the distribution of built-up areas indicated in red and purple colors by the CORINE classification (Fig. \ref{fig2}c). Since the high local image heterogeneity originates from intense backscatter plus shadow effects around vertical structures, the texture directly relates to the presence of buildings or any structure with a distinct vertical component. In the amplitude data urban areas also show high values represented by bright gray tones, but the contrast to other land cover types which also show comparably high amplitudes is not as distinct as in the texture data.

\subsubsection{Unsupervised classification}
\begin{figure}[!]
\begin{center}
\includegraphics[width=1\textwidth]{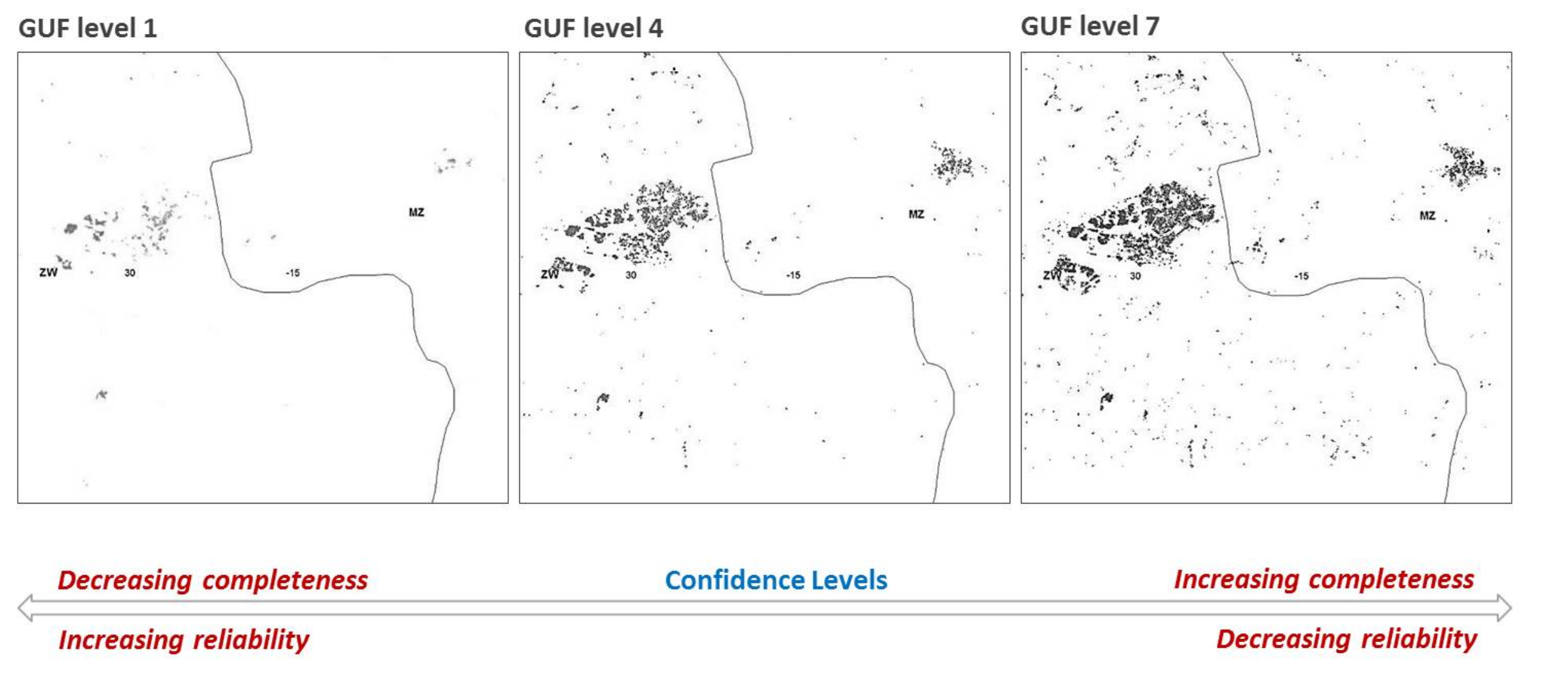}
\caption {GUF raw masks resulting from classification procedure with systematically varying confidence levels, ranging from very strict settings for level 1 and comparably relaxed parameterization for level 7.}
\label{fig3}
\end{center}
\end{figure}

The classification procedure basically couples an analysis of the backscatter amplitude data and the local texture image. For that purpose an unsupervised classification method based on advanced Support Vector Data Description (SVDD) one-class classification was implemented as described in detail in \cite{Each2013b}. For each single scene the approach identifies the optimal settings for the classification by using training samples that are automatically identified based on thresholds derived from image statistics of the amplitude and texture data, respectively. Thereby, the SVDD technique aims at i) determining the hypersphere with minimum radius enclosing all the training samples available for the class of interest - here the built-up class - and ii) finally associating all the unknown samples falling inside the boundary with it. This approach allows increasing generalization and obtaining a more consistent and reliable GUF map \cite{esch2013}. The result of the classification procedure is a binary raster layer indicating the class “built-up area” and – for all regions not assigned to this class – the category “non-built-up”. Here it is important to note that the typical characteristics of built-up areas in the amplitude and texture images are related to the location of vertical building structures. Consequently, the resulting GUF masks rather reflect the detailed building distribution and not the impervious surface that typically includes the total of buildings, roads, paved surfaces. This characteristic represents a distinct difference to settlement maps derived from (multi-)spectral satellite imagery.
The implemented classification method proved to be very robust, although over- or underestimation of built-up area might still occur when scenes show very specific land cover distributions - for instance in case of scenes covering a coastline and mostly showing water with only few areas representing land surface. Such specific land cover configurations lead to extreme distributions of the amplitude and texture statistics of the corresponding images which finally hinders the proper definition of accurate classification settings. As a result striping effects could be recognized when looking at larger collections or mosaics of GUF masks.
Comparable problems could be observed when a certain scene had been collected at a specific time or under specific environmental conditions (e.g., snow cover, soil moisture, phenological effects) that significantly differed from those of the neighboring image acquisitions. To compensate for the resulting deviations between the GUF masks, a total of six additional GUF raw versions with systematically altered classification settings - that can be considered as confidence levels - are generated. Three versions are based on constantly stricter thresholds compared to the automatically defined version (leading to the assignment of less built-up area) and three versions with more relaxed settings compared to the original definition (thus showing an increased amount of potential built-up
regions). 
As described in \cite{esch2013} an automated process is first applied to estimate a threshold for the speckle divergence image (texture layer) which is used to identify and extract a set of potential built-up areas. This set of potential built-up areas is then taken as training data sample for a SVDD one-class classification considering the texture and the amplitude image as input to assign the two categories “built-up” and “non-built-up” area. To generate the three stricter as well as the three more relaxed classification versions the speckle divergence threshold that has been defined by the automated estimation procedure is increased and decreased, respectively, three times by 200 DN with each step (e.g. if starting threshold of automatically defined version is 1800, the stricter thresholds used are 2000, 2200 and 2400 whereas the relaxed ones will take values 1600, 1400 and 1200 for the selection the training data sample). The amount of seven levels in total is selected due to technical reasons because it represents the maximum number of levels for which still all information can be stored in one single 8-bit dataset (1bit for each of the seven versions plus 1-bit for NA values) so that the data volume is kept more manageable (one global coverage in 12m includes a total of $>50$ billion pixels). Figure \ref{fig3} shows an example of the resulting GUF confidence levels. In the automated post-editing phase the availability of different GUF levels provides a flexibility in terms of built-up area representation that finally allows to effectively and optimally adapt the GUF level to the local or regional properties of the landscape (e.g., more strict levels in rough terrain where the classification approach is more prone to produce false alarms).

\subsubsection{Mosaicking}
In order to provide more manageable working units for the post-editing procedure, all the individual GUF raw masks in their original image geometry were merged to tiles of $5^{\circ}$x$5^{\circ}$ geographical latitude and longitude for each of the seven GUF confidence levels. Therewith the amount of 1,284,743 GUF masks (182,249 masks for each of the seven confidence levels) was reduced to 8309 GUF raw tiles (1187 masks for each of the seven confidence levels). The chosen tile size presents a reasonable trade-off between file size, number of tiles and practicability in terms of data and file handling. The conformity of the merge is also retained as a quality measure. 
During the mosaicking, all multiple GUF accounts in the overlapping areas of neighboring scenes were aggregated by means of a majority vote for each single pixel and each separate GUF version 1-7. As a result, each tile comprises seven GUF bands in the geometric resolution of 0.4 arc seconds ($\sim12 m$), or approaching the poles, in correspondingly lower longitudinal resolutions (e.g., north of $50^{\circ}$ N up to $60^{\circ}$ N in 0.6 arc seconds).

\subsubsection{Automated post-editing}
The highly automated post-editing stage of the GUF production is split-up into two phases: first an image segmentation is conducted that transfers all clusters of connected pixels classified as built-up in each of the seven GUF raw raster layers (confidence levels) into individual image objects. Second, a rule-based approach selects the appropriate local GUF confidence level and finally removes all GUF segments from the resulting collection that most likely represent false alarms.

\begin{itemize}
\item{ \textbf{Object generation.}}
The image segmentation and object labeling algorithm is based on the approach presented by \cite{chang2004} that uses contour tracing to iterate an image only one single time. An additional process then fills single pixel holes within the resulting objects in order to avoid negative effects of such structures to the later calculation of shape-related indices.
Concurrently, a comprehensive set of descriptive attributes is calculated for each object, including geometric properties on the one hand, and zonal statistics related to a set of reference layers on the other hand. The geometric properties include area, perimeter and the relation of area and perimeter as a measure of the object compactness.
Regarding the set of reference layers, a total of nine globally and two continentally available data collections were generated and used which are listed and further detailed in Table 1 showed in Fig.\ref{fig_table}. Basically, all reference layers except of TimeScan-ASAR (DLR-TSA) and TimeScan-Landsat (DLR-TSL) represent binary mask that were either converted from the original data formats into binary masks based on defined thresholds (applied for DLR-RM, DLR-RC, CIL) or by selecting specific classes (applied for OSM-S, OSM-R, GL30-S, GL30-W, GL30-WT, NLCD). The two TimeScan reference layers are used in their original form which is a Geotiff in float formatting  (\cite{DLR2017}). In the later, rule-based analysis phase described in section 2.2.5.2, the different binary masks are used for the optimal GUF level definition as well as for the formulation of exclusion or inclusion criteria in the context of false alarms identification.

\item{\textbf{Level selection and false-alarms removal}}
One of the key challenges for the GUF production is the quality assessment and enhancement that ensures a high accuracy and a global consistency of the final GUF data set. Generally, the input to the post-processing included the seven GUF levels separated in 8309 tiles (1187 tiles per level) and consisting of a total of more than 326 million GUF objects or 350 billion pixels, respectively. To handle this large amount of input data the post-editing was implemented in form of a highly automated methodology. The core of the post-editing module is a rule-based procedure implemented in Python that first defines for each of the 1187 tiles the optimal GUF confidence level, and second eliminates all GUF objects that are likely to not represent a human settlement. 
The selection of the best-fitting GUF confidence level is conducted per $5^{\circ}$x$5^{\circ}$ tiles by comparing the built-up area assigned in the GUF versions (1 to 7) to the built-up areas assigned by the reference layers for the same region. For that purpose, first all binary reference layers available for one tile and indicated as confirmation masks are merged by summing up the number of positive reference counts for each pixel within the $5^{\circ}$x$5^{\circ}$ tile under investigation. The reference layers are DLR-RM, DLR-RC, CIL, OSM-S, OSM-R, GL30-S, and NLCD. If two or more reference layers indicating the existence of built-up area in a pixel, such pixel is accepted and assigned a value 1. For pixels with less than two positive counts a value of 0 is assigned, meaning they are not considered to represent built-up area. Next, for each single GUF object at each confidence level (see section 2.2.3) it is calculated to what part the reference layer intersects with the GUF object. 
Starting at the most strict confidence level 1 and then proceeding up to level 7, it is then checked for each GUF object at which level its area corresponds to at least two thirds with the reference layer. The highest GUF level which still fulfills this condition is then assumed to represent the best-fitting estimation of the built-up area. Here, the basic assumption is that coming from a stage of underestimation - which can always be expected at level 1 (see section 2.2.3) - an accurate representation of the built-up area has been achieved as long as the object area is almost identical to the settlement structure provided by the reference data. For instance, since the GUF regions classified with the strict settings normally represent the most structured core zones of the settlement area, the GUF objects of confidence level 1 or 2 usually show an almost $100\%$ proportion of reference data. However, the error of omission – meaning correct settlement regions that have not yet been assigned by the classification – is still comparably high at these lower levels. Moving upwards to the higher GUF confidence levels, the effect of underestimation then constantly decreases while the probability of overestimations increases (see\ref{fig3}). We therefore expect overestimation to start as soon as less than two thirds ($< 66\%$) of the object area corresponds with the reference data. The threshold of $66\%$ was defined by empirical tests on the basis of local ground truth data available for various globally distributed test regions.
The final step of the optimal confidence level definition includes a statistical analysis (majority vote) that defines the GUF level which has most frequently been assigned to the objects within the 5°x5° tile when applying the $66\%$-rule described before. The corresponding level and its objects, respectively, are then used as the optimal GUF version for the tile. In case that the reference information is just available for less than $5\%$ of all GUF objects within a tile, it is assumed that the selection procedure cannot be expected to generate a reliable outcome. Hence, the GUF level 4, representing the primary outcome of the GUF classification procedure (see section 2.2.3), is chosen.
The final step of the post-editing phase includes a procedure to identify and eliminate false alarms that might still be present in the previously selected best-fitting GUF layer. For that purpose the exclusion masks DLR-RM (where layer "shaded relief " $>212$ or where layer "roughness" $>15$), GL30-W (where layer "GL30" = 50), GL30-WT (where layer "GL30" = 60) as well as the reference layers DLR-TSA and DLR-TSL and the merged confirmation mask (MCF) used for the optimal level selection are used in combination with a simple rule-based analysis. Thereby, in a first step all GUF objects are deleted if they overlap to more than one third ($33\%$) of their area with water bodies (GL30-W), wetlands (GL30-WL) or the relief mask (DLR-RM) and if they don't show at least a proportion of $>33\%$ of built-up area. Here, the built-up area is indicated by the MCF already used for the optimal level selection. Moreover, objects are erased if their zonal statistics related to the two TimeScan layers DLR-TSA and DLR-TSL indicate an erroneous assignment as built-up area. This is the case when objects show a high value for the mean temporal NDVI ($> 0.6$) or a low value for the mean temporal backscatter (sigma nought $< 0.1$). Typically, further positive false alarms, e.g. related to highly textured forests (here the temporal NDVI is too high) or originating from highly textured rice field areas (here the temporal SAR backscatter is too low), can be successfully identified and eliminated with this procedure. An exemplary result for the rule-based false alarms identification and elimination is shown in Figure \ref{fig4}. Summarizing, the rule set applied to each GUF candidate object removes the according object i) if proportion of MCF $<30\%$ and proportion of GL30-W $>30\%$ or proportion of GL30-WT $>30\%$ or proportion of DLR-RM $>30\%$, or ii) in case that the zonal mean of mean temporal NDVI from DLR-TSL $>0.6$ or the zonal mean of mean temporal sigma nought from DLR-TSA $<0.1$. Generally, it is important to note that no procedure has been implemented to correct errors of omission in order to assure consistency of the GUF data since any other decision would mean that the final result might include areas - or geometries - that don't originate from the original SAR data and classification procedure, respectively.

\end{itemize}

\begin{figure}[!]
\begin{center}
\includegraphics[width=0.6\textwidth]{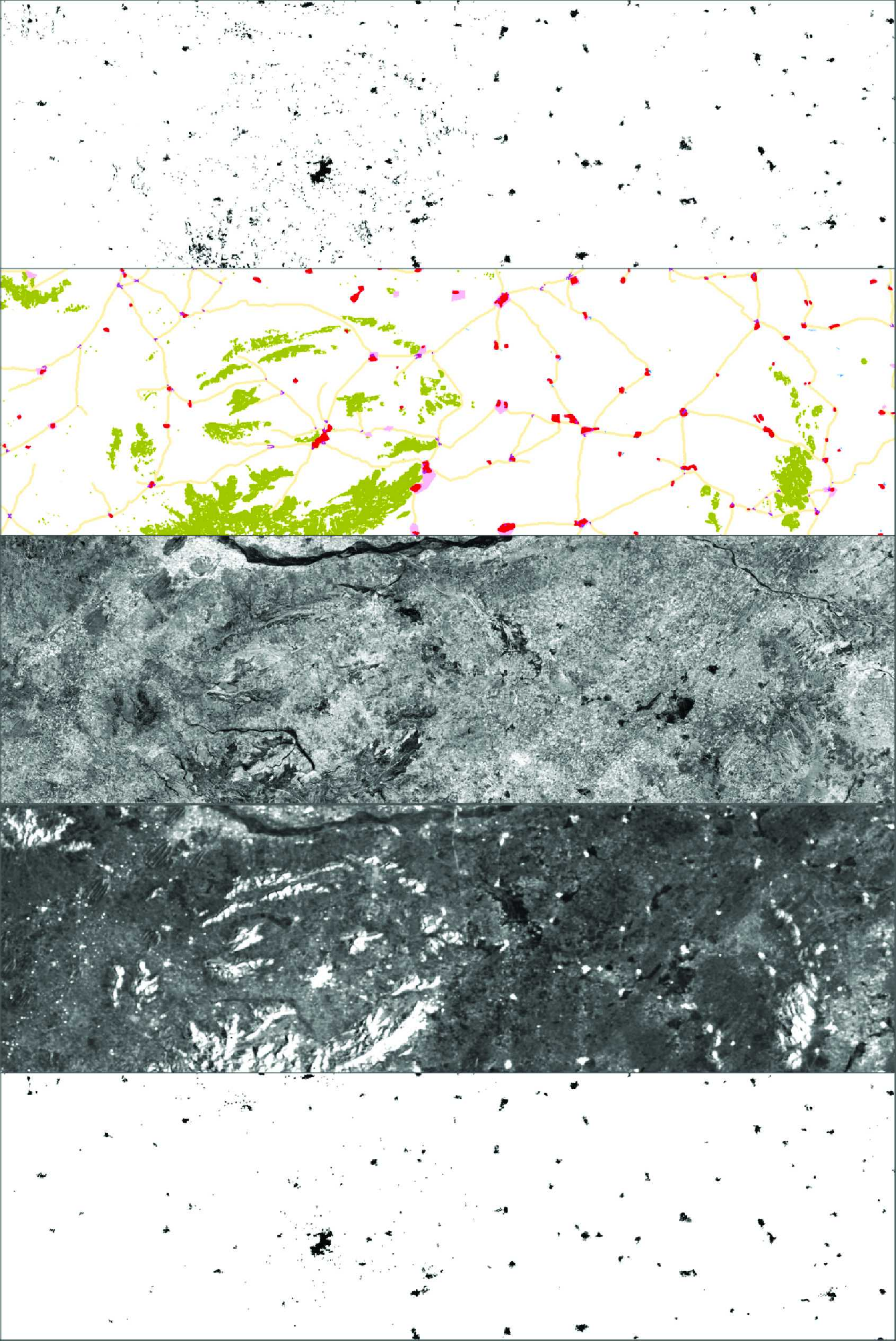}
\caption {Example of identification and removal of false positive alarms based on automated post-editing procedure with selected optimal GUF level (top), selection of thematic reference layers including GL30-S (red), GL30-W (blue), OSM-S (rose), OSM-R (orange) and DLR-RM (green), then the TimeScan-Landsat (temporal mean of NDVI)  and TimeScan-ASAR (temporal mean of sigma nought) reference data, and the corrected final GUF product (bottom).}
\label{fig4}
\end{center}
\end{figure}


\begin{table}[!]

\caption {Binary reference layers used in the context of the post-editing procedure. Data from: \cite{GDAL,DLR2017,NASA2012,NASA2016a,NASA2016b,Kennelly, zhou1992,OSM,Chen20157,EEA,COPERNICUS,MRLC}}
\label{fig_table}
\tiny{
\begin{tabular}{|p{2.5cm}|p{4cm}|p{2.5cm}|p{1.8cm}|c|}
\hline
\bf Reference Layer  & \bf Description & \bf Source(s) & \bf Mask type & \bf Coverage\\ \hline 
Relief mask - DLR-RM & Mask generated based on DEM shaded relief (modelled on the basis of mean incidence angle and orientation regarding SAR line of sight) combined with DEM roughness; both derived from SRTM version 2.1 DTM in 3 arcsec and ASTER GDEM V2  & \cite{GDAL, NASA2016a, NASA2016b, Kennelly,zhou1992} & Exclusion: indicating areas with critical relief (IF "shaded relief  $>212$" OR "roughness $>15$" then "relief mask = 1") & Global \\ \hline
OSM-Settlements - OSM-S & Settlement-related categories extracted from Open Street Map point and polygon data & \multirow{3}{*}{\cite{OSM}} & Confirmation: indicating built-up areas & \multirow{3}{*}{(Global)}\\ 
OSM-Roads - OSM-R & Road-related categories extracted from Open Street Map line data & & Confirmation: indicating built-up areas & \\ 
DLR Road Clusters - DLR-RC & Self-made data set indicating local clusters of roads by applying focal mean filter to binary road network layer derived from Open Street Map data & & Confirmation: indicating concentrations of roads &  \\ \hline
 GL30-Settlements - GL30-S & Artificial surfaces (class 80) & \multirow{3}{*}{\cite{Chen20157}} & Confirmation:  indicating built-up areas & \multirow{3}{*}{Global}\\ 
 GL30-Water - GL30-W & Water (class 50) & & Exclusion: indicating un-inhabited water areas (if "GL30 = 50" then "GL-30-W = 1")
 & \\ 
 GL30-Wetlands - GL30-WL & Wetlands (class 60) & & Exclusion: indicating un-inhabited wetland areas (if "GL30 = 60" then "GL-30-WL = 1") & \\ \hline
 Copernicus Imperviousness Layer 2012 - CIL & Extraction of all areas with an imperviousness of GT 30\% & \cite{EEA, COPERNICUS}& Confirmation: indicating built-up areas & Europe \\ \hline
 US National Land Cover Dataset 2011 - NLCD & Classes 22, 23 and 24 from category Developed & \cite{MRLC} & Confirmation: indicating built-up areas & US\\ \hline
 TimeScan-ASAR - DLR-TSA & Self-made long-term SAR backscatter data derived from Envisat-ASAR WSM data collected from 2010-2012 (25,550 scenes) & \cite{DLR2017} &  Exclusion: indicating areas with constantly low backscatter (if "Mean sigma nought of TimeScan-ASAR  $<0.1$" then "DLR-TSA = 1") & Global\\ \hline
 TimeScan-Landsat - DLR-TSL & Self-made long-term mean NDVI data set derived from Landsat-5/-8 data collected from 2013-2015 (418,918 scenes) & \cite{DLR2017} & Exclusion: indicating areas with constantly high NDVI (if "Mean-NDVI of TimeScan-Landsat  $>0.6$" then "DLR-TSL = 1")& Global\\ \hline

 \end{tabular}
}

\end{table}

\section{The new Global Urban Footprint data set}
Due to the TanDEM-X PPP data policy (see section 2.1) any commercial use of the GUF product is subject to a licensing by Airbus DS, whereas DLR grants free and open access to the full-resolution (0.4 arcsec) GUF global dataset for any scientific use. For any non-profit, but also non-scientific use scenario, DLR and Airbus DS agreed upon a free and open access to a spatially generalized GUF version in 2.8 arcsec spatial resolution ($\sim84m$).

\subsection{Product specification}
\begin{figure}[!]
\begin{center}
\includegraphics[width=1\textwidth]{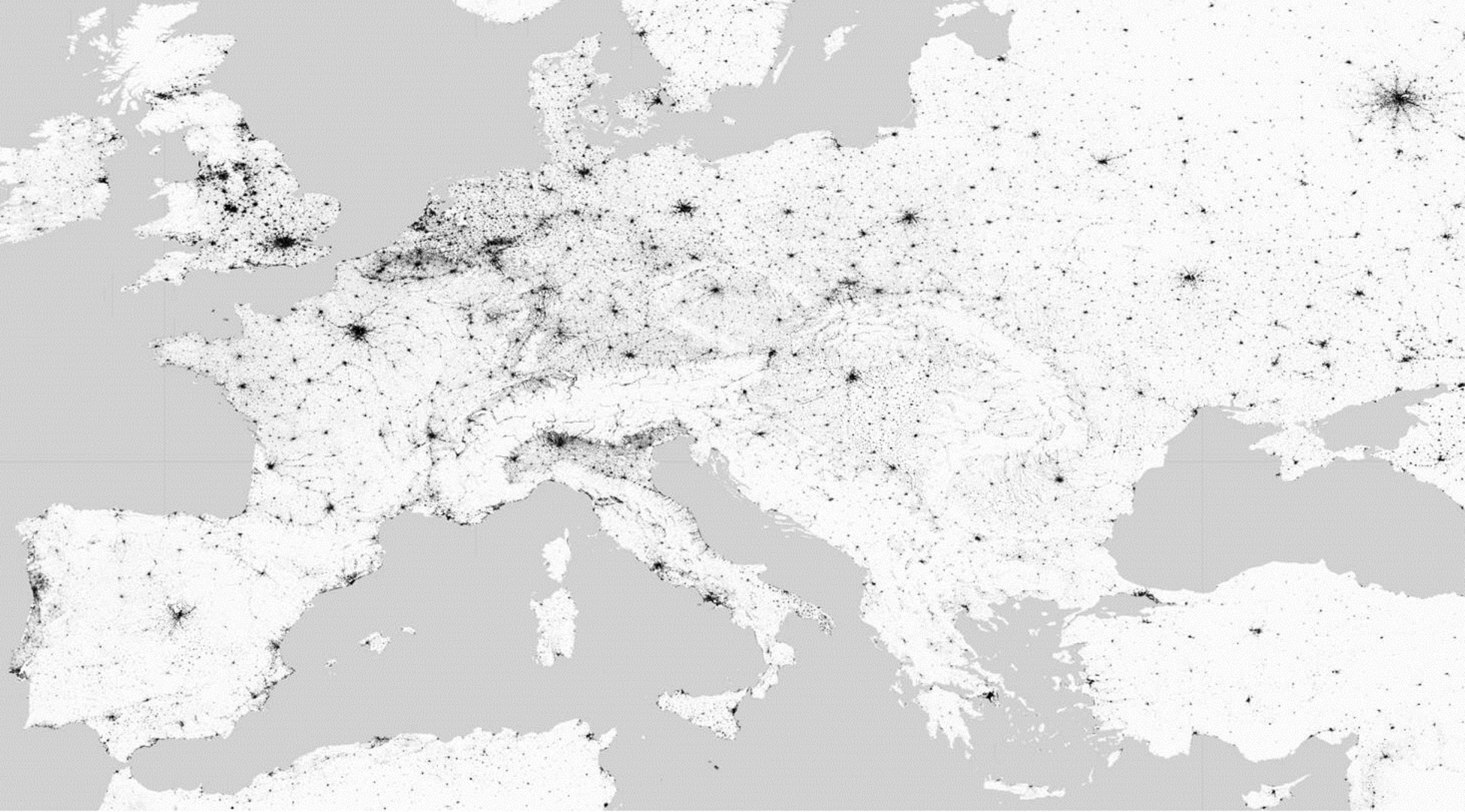}
\caption {Overview of GUF product for Europe.}
\label{GUF}
\end{center}
\end{figure}

\begin{figure}[!]
\begin{center}
\includegraphics[width=0.6\textwidth]{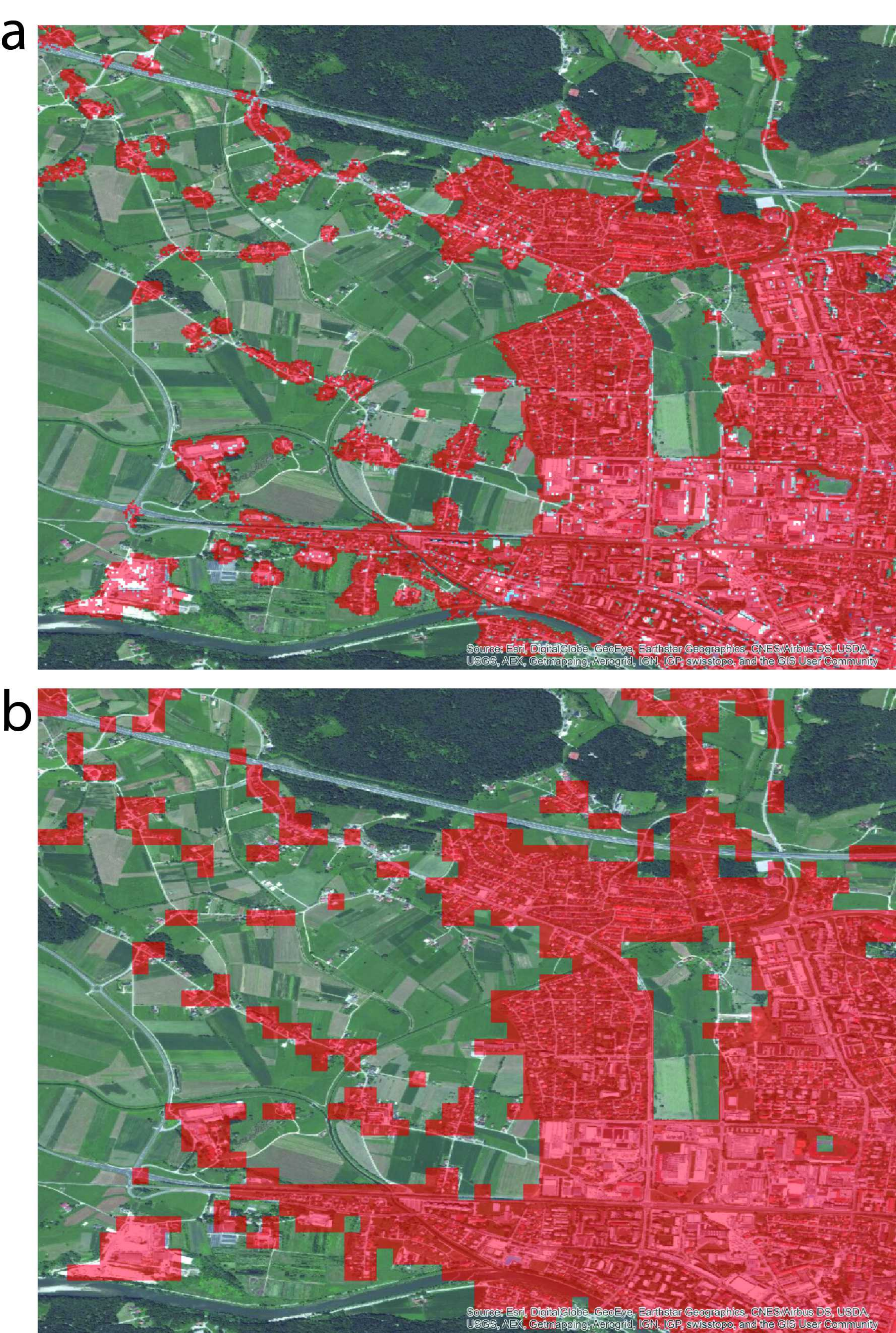}
\caption {Comparison of spatial detail provided by original GUF product in 0.4 arcsec geometric resolution (a) and the generalized version in 2.8 arcsec (b), in red the classified built-up ares.}
\label{fig6}
\end{center}
\end{figure}

This sub-section details the basic product specifications of the full-resolution GUF layer in 0.4 arcsec geometric resolution and the version for non-profit/non-scientific use in 2.8 arcsec. Fig.\ref{GUF} presents a  subset of the global GUF for Europe. The GUF layer is provided as binary raster data sets in 8-bit, LZW-compressed GeoTiff format with a value of 255 indicating built-up area, a value of 0 representing all non-built-up areas, and no data assigned by the value 128. The projection is Geographic coordinates (Lat, Lon). The geometric resolution of the original GUF data is 0.4 arc seconds which corresponds to 12 m near the equator while the GUF layer for any non-scientific/non-commercial use shows a reduced resolution of 2.8 arc seconds (84 m near the equator). Towards the poles the spatial resolution decreases to 0.6 arcsec between $50^{\circ}$-$60^{\circ}$ N/S, 0.8 arcsec from $60^{\circ}$-$70^{\circ}$ N/S, and 1.2 arcsec in areas $>80^{\circ}$ N/S.
The generalized GUF version in 2.8 arcsec is directly derived from the 0.4 arcsec version by assigning a value of 255 (= built-up) to all pixels whose coverage contains a proportion of $>25\%$ GUF area as defined by the original 0.4 arcsec data. Hence, the spatially reduced version shows a level of detail which is almost identical with the original resolution, as shown in Fig.\ref{fig6}.

\subsection{Quality assessment and comparison with existing EO-based settlements layers}

\begin{figure}[!]
\begin{center}
\includegraphics[width=0.9\textwidth]{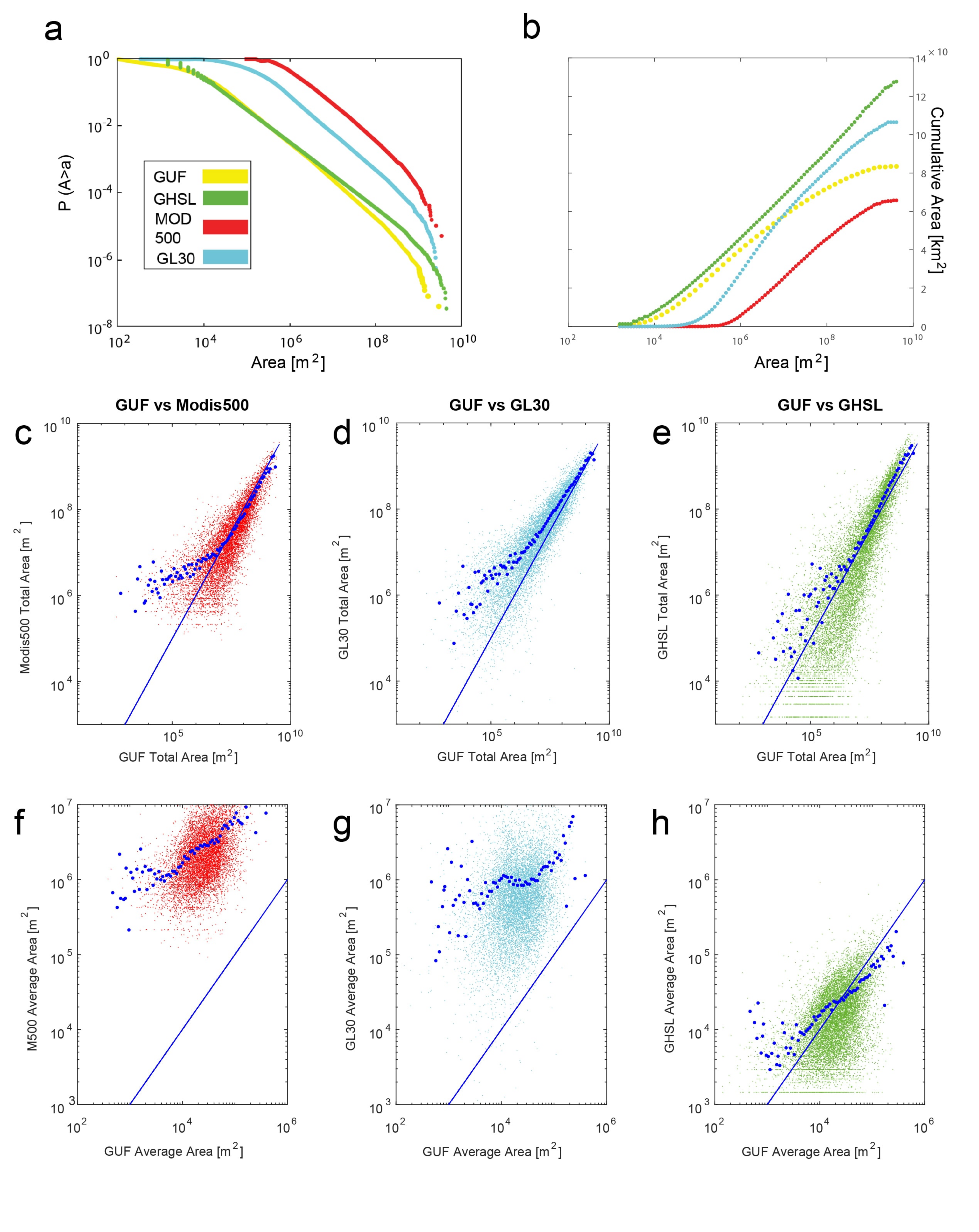}
\caption {Comparison of global urban inventories. All colors in the figure are consistent with the legend in (a). (a) cumulative distribution function of the areas of all built-up objects. (b) cumulative plot of the areas.  (c,d,e,f,g,h) one degree tiles summary statistic plots. (c,d,e) show the plot of the total area of classified object in each tiles for GUF vs MOD500 (c), GUF vs GL30 (d) and GUF vs GHSL (e). Blue dots in c,d and e are the average total area of tiles in a log-binned interval of the GUF areas. In the same manner in f,g and h the plot is for the average area for each one degree tiles.}
\label{fig_stat}
\end{center}
\end{figure}

\begin{figure}[!]
\begin{center}
\includegraphics[width=1\textwidth]{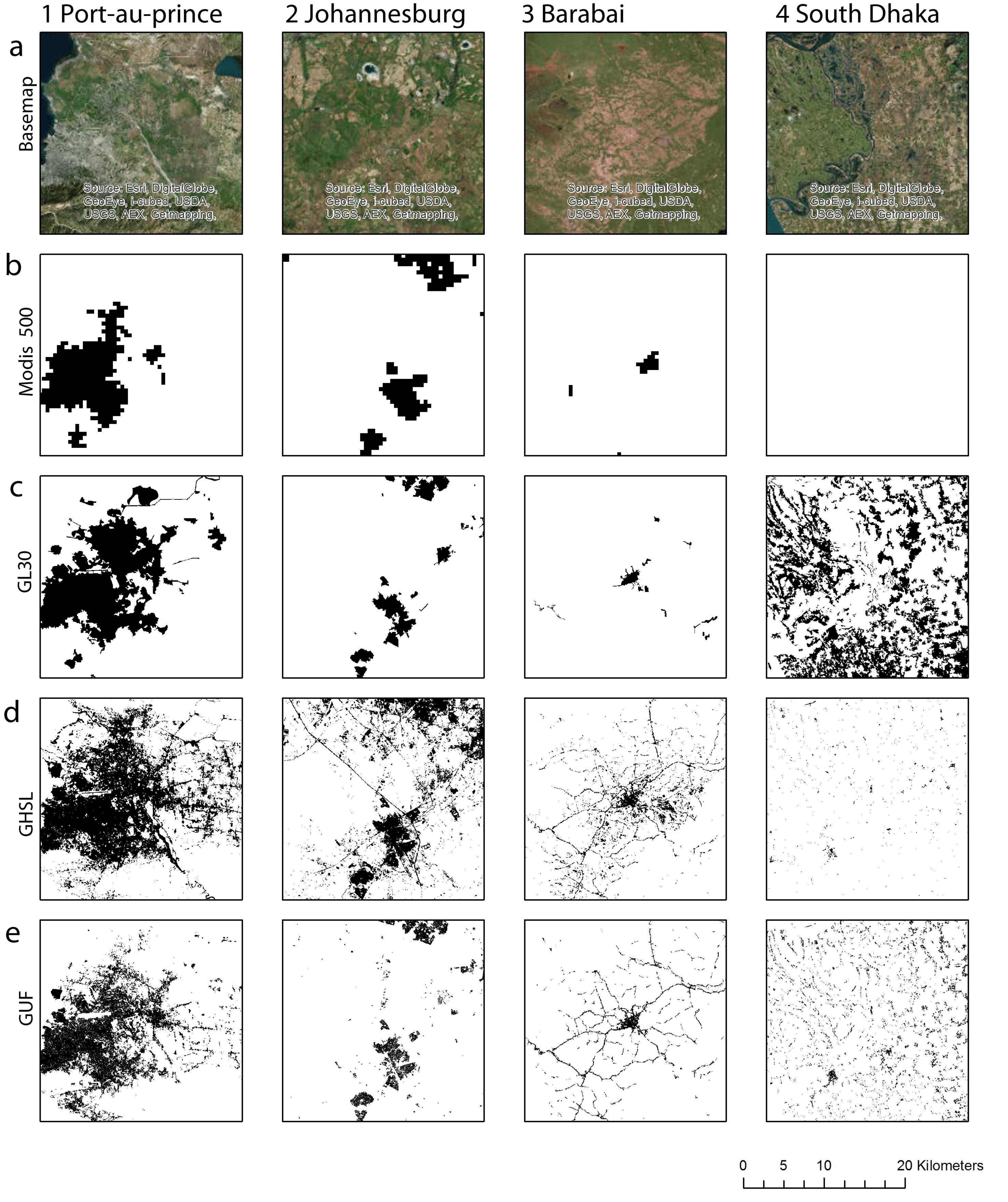}
\caption {Samples extracted from the four global urban inventories. }
\label{fig_cases}
\end{center}
\end{figure}

On regional scale, the thematic accuracy and the strengths and weaknesses of the GUF layer have already been documented in detail by comparisons to ground truth data in several studies and projects such as presented by \cite{Each2013b},  \cite{Felbier2014}, and  \cite{Gessner2015282} .  \cite{Klotz2016191} conducted a detailed multiscale cross-comparison between the GUF layer and existing low (MOD500, GlobCover) and high resolution (GHSL-SPOT2.5m) human settlements data derived from EO imagery. The study included an assessment of relative inter-map agreement, absolute accuracies and pattern-based classification differences. The results showed that the GUF data holds particular advantages regarding the representation of small-scale rural settlement patterns which are significantly underestimated in the lower resolution products. Here, the Kappa coefficient of agreement indicated values between $0.56$ - $0.58$ for the GUF which was almost twice as good as the Kappa values for MOD500 or GlobCover. At the same time the assessment proved consistent accuracies between urban (K = $0.46$-$0.50$) and rural landscapes (K = $0.41$-$0.45$). Thereby, the pattern-based analyses indicated a significant correlation of the observed accuracies and the settlement density, with generally decreasing accuracies from large to medium and small patch sizes. Nevertheless, it could be demonstrated that even quite small patches are frequently identified, which are completely missing in the low resolution maps.
Regarding the limitations of the GUF layer, the outcomes of the detailed local and regional investigations showed that false positives (overestimation) can occur in landscapes where the basic physical structures are quite similar to those of settlements - e.g., ensembles of large rocks, steep embankments of rivers, mineral extraction sites, or regularly dispersed groups of big trees. High local textures and backscattering values can also occur in the context of rice fields or wetlands mixed with sparse vegetation. It could also be observed that local patterns of increased backscatter and texture levels were identified in specific collections over rain-forest regions, whereas there are indications that the amount of false alarms is related to increased surface wetness/moisture originating from intense rainfall events. Errors of omission occur when built-up areas are composed of houses with flat roofs that are located very close to each other so that - in fact – the walls are not illuminated by the radar beam so that – in effect - the typical double bounce reflection does not appear. The same effect develops when buildings are surrounded by high trees so that only the roofs are visible but not the walls. Also situations when buildings are oriented at around $45^{\circ}$ to the incoming radar beam can cause problems in terms of a proper recognition since in these cases the incoming radiation is mostly reflect away from the sensor. \\
In this section we also provide a comparison of GUF versus other widely used inventories of human settlements at global scale, namely MOD500, GL30 and GHSL. We compared the four data sets in form of i) the distribution of the area of all classified urban objects at global scale and ii) summary statistic of objects' area in $1^{\circ}$x$1^{\circ}$ tiles.  Results are presented in Fig.\ref{fig_stat}. In order to interpret the results presented in Fig.\ref{fig_stat} we also provide examples of  four demonstration sites in Fig \ref{fig_cases}, which represent different urbanization typologies for different regions and area range. Here it has to be stated that none of the layers can be considered as ground truth. So this comparison gives only information on the level of agreement between the GUF layer and the other data sets on built-up areas. 

A systematic global campaign to assess the absolute accuracy of the new GUF product compared to ground truth data is currently ongoing, but will still take several months to be finished. Nevertheless, we believe that the outcomes of the a relative cross-comparison presented in this paper can already be of high interest for users and stakeholders of the corresponding data sets.

 In Fig.\ref{fig_stat}a we plotted the cumulative distribution function for the areas of all classified objects for each data set. The slope of such distributions, coherent with the typical power-law or Zipf's distribution (\cite{Zipf,clauset2009}), is remarkably similar between GUF, MOD500 and GL30, while for GHSL the curve is less steep - indicating an abundance of large classified objects specially if compared to GUF which presents similar maximum size. 
At the top left of the plots provided by Fig.\ref{fig_stat}a the different starting points of the curves indicate different minimum sizes of the built-up objects detected - an effect that basically originates from the different spatial resolutions of the data. One can observe that the GUF curve starts with the lowest object sizes compared to the other data sets. Differences are also emerging considering the total built-up areas indicated by each data set with GUF 834,260 $km^2$, GHSL 1,289,506	$km^2$, GL30 1,064,940 $km^2$ and M500 65,7384 $km^2$.

Fig.\ref{fig_stat}b plots the total accumulated area of all settlements against the size of the settlement objects. We can observe here that the total area for GHSL is constantly larger over the entire range of object sizes with respect to the other layers. The MOD500 data set, in turn, constantly shows the lowest values. Compared to the GUF, the GL30 assigns significantly less settlement area for the lower settlement sizes whereas for the larger objects it shows way more built-up regions. This effect might arise from the lower spatial resolution in combination with a relative overestimation due to the inclusion of roads, open paved places etc. which are not mapped by the GUF (see section 2.2.3). In terms of total urbanized area GHSL is 1.5 times larger than GUF. GUF and GL30 intersect at $A\sim10^7 [m^2]$ with GUF exceeding GL30 for objects of area $<10^7 [m^2]$ while for larger areas GL30 assigns more built-up with a final total area which is 1.27 times larger than this indicated by the GUF. The constantly lower values for MOD500 arise from the significantly lower spatial resolution compared to all other layers.
Based on $1^{\circ}$x$1^{\circ}$ tiles we also analyzed the agreement of the overall sum of the areas of the mapped settlements objects and their average sizes. In total, considering only tiles with presence of built-up areas in the GUF, we used 29240 tiles. Fig \ref{fig_stat}c,d,e show the corresponding plots with the values for GUF on the x-axes and MOD500 (c, red), GL30(d, blue) and GHSL(e, green) on the y-axes. The blue points in the plots indicate the average total settlement area with each point representing the value calculated for one of 100 log-binned ranges defined for the entire range of areas occurring in the data. The blue lines correspond to the linear function (bisector) of GUF points. Analogous, in Fig \ref{fig_stat}f,g,h, the average area of settlement objects in each tile is shown by corresponding graphs. Basically, all of these plots confirm great differences between the existing inventories - a view which is confirmed by the examples

shown in  Fig \ref{fig_cases}. For example MOD500, in the area range $>10^6$ to $>10^{10}$, the sum of the areas of classified objects is always lower then GUF (Fig.\ref{fig_stat}c). Such differences can be explained by the absence of small settlements in MOD500 as is visible also in Fig.\ref{fig_stat}f and Fig \ref{fig_cases}.
In Fig.\ref{fig_stat}d,e we see that GHSL and GL30 always take greater values than GUF over the entire spectrum. But while for GL30 the sum of areas in each tiles converges to  the GUF areas (Fig.\ref{fig_stat}d), GHSL keeps diverging beyond $A=10^7$ as well, meaning that the bigger the sum of areas of settlement objects the larger the differences between GUF and GHSL. For GHSL that might be due to a scaling effect of over estimation or by the abundances of smaller and scattered settlements in GHSL under a certain size threshold (see Fig.\ref{fig_cases}d1, d2, d3). This effect is consistent with Fig.\ref{fig_stat}h where we observe a double trend in comparison of the mean sizes.
Great divergences in both total classified size and statistic over tiles warrant a more systematic comparison between present data bases, which is  is currently ongoing and out of the scope of the present study.

\section{Conclusions and outlook}
With the new Global Urban Footprint (GUF) layer the German Aerospace Center (DLR) has generated the spatially most detailed and consistent global map of human settlements so far. With a 3m geometric resolution of the input satellite imagery and 12m pixel spacing for the resulting human settlements map, the GUF data represents a precise inventory of both large urban agglomerations as well as the dispersed small-scale built-up areas in rural regions. Therewith, the GUF facilitates detailed quantitative and qualitative analyses and comparisons of settlement properties and patterns from the local municipal level up to the global scale. In this context the GUF layer also benefits from the fact that the global input data could be collected within just 2 years. Due to the SAR imaging principle the GUF is at the same time not affected by any disturbing atmospheric effects such as cloud coverage or differing sun illumination – a significant advantage compared to optical multispectral data collections. Hence, the GUF data base shows a unique spatio-temporal consistency that predestines it as a baseline layer or starting point for comparative global and/or multi-temporal studies. The SAR imaging principle also comes along with another positive effect: while optical multispectral imagery rather reflects the chemical properties of the observed surfaces in form of a spectral signature, the SAR backscatter is basically influenced by the structural or geometric features of the objects. And these physical/structural properties of settlements are more similar all over the globe since almost any typical man-made building structure shows vertical walls that finally lead to the characteristic appearance of built-up areas in SAR imagery. In contrary, the building or roof material which defines the appearance in multispectral data largely differs all over the world. Here it is important to note that in the SAR-based GUF approach the representation of built-up area might significantly differ from the definition of built-up area derived from optical data. Maps from optical data basically reflect the impervious surface – meaning that besides buildings also roads, parking lots, runways of airports, etc are included. These structures or features are not assigned as built-up areas in the GUF since they don't show any vertical component leading to the required local texture. 
With the described characteristics the new GUF dataset can help to acquire a better understanding of the urbanization phenomenon and to respond appropriately to future challenges related to sprawling cities, population explosion, poverty reduction, economic growth, climate change and carbon emissions, and the loss to biodiversity. The GUF provides uniform data – applicable worldwide – on the location, size and shape of settlements. This is a crucial advantage, especially in remote and underdeveloped regions of the Earth, where suitable geographical data are frequently scarce. This characteristic makes the GUF data very valuable to a broad spectrum of user communities from science, development organizations, non-governmental and governmental institutions, and the business sector.
DLR has released the global GUF data set to be used open and free of charge at full spatial resolution for any scientific use, and at a generalized resolution of about 84m for any non-profit applications. Currently available by email request from DLR (guf@dlr.de), the data sets are also accessible on the European Space Agency's Urban Thematic Exploitation Platform (via: https://urban-tep.eo.esa.int/) and at DLR-EOC's geoservice (via: https://geoservice.dlr.de/web/maps/eoc:guf:4326). The Urban Thematic Exploitation Platform (U-TEP) is supposed to initiate a step change in the use of EO data by providing an open and participatory platform based on modern information and communication technologies and services that enables any interested user to easily exploit and generate thematic information on the status and development of the built environment\cite{Esch2016b}. 
Regarding the future activities, DLR has already started several studies and collaborative initiatives to create new knowledge on the built environment that can help answering key questions related to global urbanization: How many settlements are on Earth and which proportion of the land surface is covered by built-up area? What is the balance between urban and rural settlements? Which relations exist between the human settlements pattern and basic natural, economic and social parameters? 
Moreover, DLR plans to release regular updates of the GUF layer based on Sentinel-1/-2 and Landsat data. At the same time there will be an extended suite of GUF+ products covering further thematic detail compared to the current binary GUF mask. This includes a GUF-NetS layer generated on the basis of a spatial network analysis (\cite{Esch2014212}) and describing the global settlements pattern and properties. A GUF-DenS layer will further specify the built-up density and urban green within those areas assigned as settlements by the conventional GUF based on a combination of the GUF mask and imperviousness/greenness information derived from TimeScan data (\cite{DLR2017}). Another product of the GUF+ suite is the GUF-TrendS product that aims at the provision of information of the spatiotemporal development of the human settlements based on the analysis of Landsat and Envisat-ASAR archive data. Finally, the generation of a GUF-VolumeS layer is planned which estimated the average volume of the built-up areas from TanDEM-X DEM data based on a procedure introduced by \cite{marconcini2014}.

\section{Acknowledgements}
\label{5}
The authors would like to thank the TerraSAR-X and TanDEM-X Science Teams for providing the global SAR data used to derive the GUF layer. We also thank the World Bank and the Swiss State Secretariat for Economic Affairs (SECO) for their financial support in the context of the GUF quality assessment and enhancement activities (Contract 7174285). The authors would also like to acknowledge the Center for International Earth Science Information Network of the Columbia University - USA (K. MacManus), the Comisión Nacional para el Conocimiento y Uso de la Biodiversidad – Mexico (R. Ressel, M. Schmidt), the Natural Resources Canada – Canada (Z. Ying), the South African National Space Agency (N. Mudau, P. Mangara), the TERI University – India (P. Joshi, R. Sharma), the University of Oklahoma – USA (Y. Hong), the University Putra Malaysia (H. Z. M. Shafri), and the University of Tsukuba – Japan (F. Yang) for the provision of local ground truth or reference data, and ENAC(EPFL) for its contribution. The authors also thank the European Space Agency (ESA) for funding the project “Urban Thematic Exploitation Platform – TEP Urban” (ESRIN/Contract No.4000113707/15/I-NB) since the processing of the global TimeScan product based on Landsat-8 data could be realized in the context of this initiative.Finally, the authors credit to ArcGIS World Imagery with the sources Esri, DigitalGlobe, GeoEye, Earthstar Geographics, CNES/Airbus DS, USDA, USGS, AEX, Getmapping, Aerogrid, IGN, IGP, swisstopo, and the GIS User Community, for the use of basemap data for the visual comparisons in Figure 5. E.S. has been founded by Swiss National Science Foundation, grant N.P2ELP2-165150.

S\section{Bibliography}
 \bibliography{bib}

\end{document}